# The Outer Regions of the Galactic Bulge. I Observations

Rodrigo A. Ibata[1,2] and Gerard F. Gilmore[1]
1 *Institute of Astronomy, Madingley Road, Cambridge CB3 0HA*
2 *Department of Geophysics and Astronomy, University of British Columbia, Vancouver, Canada (present address)*

**ABSTRACT**
An observational survey of stars selected from the region of sky in the direction of the Galactic bulge is presented. We discuss the choice of tracer populations for this study. Digitised UK Schmidt plate photometry, calibrated with CCD photometry, is obtained in 18 regions over the bulge. Stars are selected for spectroscopic follow-up from within a carefully chosen colour-magnitude selection window, optimised for efficient detection of bulge K giants. Some 1500 stellar spectra were obtained, with the AAT AUTOFIB facility, giving the largest data set yet obtained in this region of the Galaxy, by a factor of $\approx 10$. We have derived a radial velocity and metallicity for each star, and have quantified the uncertainties in these measurements. Luminosity classification has been derived both by visual classification and using an automated routine based on Principal Component Analysis. There are two basic results from this survey: the discovery of the Sagittarius dwarf galaxy, which is described by Ibata, Gilmore & Irwin (1994,1995), and a study of the Galactic Bulge, which is presented in an accompanying paper (Ibata & Gilmore 1995).

## 1 SELECTION OF THE STELLAR SAMPLE

### 1.1 Selection of Galactic Regions

In this study we aim to find stars that trace the kinematic and abundance structure of the inner parts of the Galaxy. Eighteen fields were selected for investigation from low reddening areas in the Burstein & Heiles (1982) reddening maps; these are situated approximately at the intersections of $\ell=(-25°, -15°, -5°, 5°, 15°, 25°)$ and $b=(-12°, -15°, -20°)$. The fields span most of one hemisphere of the the galactic bulge, excluding for the very central core, and also a significant part of the halo: if the major axis scale length in the outer parts of the bulge is $\approx 0.9$ kpc (*cf* Ibata & Gilmore 1994 – Paper II), which is similar to the halo scale length ($\approx 1.1$ kpc $= r_e/(1+\sqrt{2})$, where $r_e = 2.7$ kpc — de Vaucouleurs and Pence (1978), the above grid covers about four bulge or halo scale lengths on either side of the minor axis, while if bulge and halo flattening is $\approx 0.6$ (*cf* Paper II), then the grid also covers a range from about two to four bulge or halo scale lengths south of the galactic equator. We emphasise that this study is related to the outer bulge: the central core is that which most studies of the Bulge have addressed to date. In that sense we complement such studies.

The Schmidt plates that cover these regions were digitised (section 2.1) and calibrated with CCD photometry (section 2.4). The resulting photographic photometry allows desired tracer objects to be selected by colour and magnitude (section 1.3).

Tracer object spectra were gathered in six of the above 18 regions, located at $\ell, b = (-25°, -12°), (-15°, -12°), (-5°, -12°), (5°, -12°), (5°, -15°), (5°, -20°)$. These fields form what are effectively two long-slits across the sky, one 30° long and parallel to the bulge major axis, the other 8° long and perpendicular to the bulge major axis. The stars selected for spectroscopic observation were chosen from high surface brightness patches on the digitised Schmidt plates: the criterion used was that the local mean intensity had to be 2$\sigma$ over the plate mean intensity. These high surface brightness patches are expected to correspond to the least obscured lines of sight to the bulge.

**TABLE 1:** The mean reddenings in the fields investigated.

| Field $(\ell, b)$ | $E_{B-V}$ | Field $(\ell, b)$ | $E_{B-V}$ | Field $(\ell, b)$ | $E_{B-V}$ |
|---|---|---|---|---|---|
| $(-25°, -12°)$ | 0.12 | $(-25°, -15°)$ | 0.10 | $(-25°, -20°)$ | 0.09 |
| $(-15°, -12°)$ | 0.17 | $(-15°, -15°)$ | 0.09 | $(-15°, -20°)$ | 0.06 |
| $(-5°, -12°)$ | 0.15 | $(-5°, -15°)$ | 0.10 | $(-5°, -20°)$ | 0.08 |
| $(5°, -12°)$ | 0.15 | $(5°, -15°)$ | 0.12 | $(5°, -20°)$ | 0.08 |
| $(15°, -12°)$ | 0.21 | $(15°, -15°)$ | 0.18 | $(15°, -20°)$ | 0.15 |
| $(25°, -12°)$ | 0.27 | $(25°, -15°)$ | 0.21 | $(25°, -20°)$ | 0.15 |

### 1.2 Selection of the Tracer Population

While most of the stellar mass in the Galaxy is in very low mass stars, it is not necessary to study such sources to obtain a fair tracer of the inner Galaxy. In addition, acquisition of spectra for such stars at the distance of the Galactic bulge would require unrealistic amounts of large telescope time. An ideal tracer is the population of K giants, for the following reasons:

(i) The lowest bulge K giant surface density along any of the lines of sight investigated is $\approx 1300$ stars deg$^2$. The

multi-object spectrograph available to this study (*cf* §3) can simultaneously observe $\approx 50$ objects over a field of $0.35\,\mathrm{deg}^2$: thus K stars have sufficiently high surface density to make efficient use of the technology.

(ii) K giants are bright (*cf* §1.3), so it is possible to obtain a large sample and hence good statistics with minimal telescope time.

(iii) Since it can be ambiguous from a star's colour and apparent magnitude whether it belongs to a distant intrinsically luminous population or an intrinsically faint nearby population, the selected tracer population should have spectra that allow this luminosity discrimination. K giants can be discriminated from K dwarfs with low resolution spectra reliably. This discrimination is less reliable for M stars from the $2\,\mathrm{\AA}$ resolution spectra obtained in this investigation, while the use of F and G stars would have required precise luminosity classification to remove subgiant contamination, which is difficult to obtain.

(iv) K giant spectra have features primarily dependent, in decreasing order of importance, on effective temperature, abundance and surface gravity. It is possible to calibrate this dependance on physical parameters both empirically (Faber et al.1985, Friel 1986) and by creating theoretical stellar atmosphere models (Cayrel et al.1991). K giant abundances can be estimated to $\approx 0.25\,\mathrm{dex}$ (Friel 1986) if their colour is known and if they can be distinguished from dwarfs. However, abundance determinations for cooler stars from the data of the type available here are more problematic.

(v) The selection criteria are well understood, as all stars pass through the K giant phase. This gives a fair sample of the underlying stellar distribution.

The reddest few stars in each field were also observed, in order to limit the population of bulge carbon stars. Five carbon stars were found (*cf* Ibata, Gilmore & Irwin 1994) from the $\approx 30$ very red stars ($\approx 2\%$ of the full sample) observed. The criteria used for spectral-type classification, including carbon star - M giant discrimination and K dwarf-K giant discrimination, are described further in §8.

### 1.3 Colour-Magnitude Selection of Stars for Spectroscopic Observation

Stars were chosen for spectroscopic observation from the digitised photographic photometry by imposing the following colour-magnitude (CM) criteria: $(B-V)_0 \gtrsim 1.0$ and $14.5 \lesssim V \lesssim 16.5$. The CM cuts are actually performed on $(B_J - R)$ and R photometry (*cf* section 2.4 for the colour conversions) and are slightly different in each field because the line-of-sight tangent-points have slightly different distance moduli (at most $0.2^\mathrm{m}$) and because the reddenings (listed in table 1) are slightly different in each field. A justification for this CM selection window follows.

The galactic regions investigated are all situated at relatively low latitudes, where disk stars are the most numerous population. Since this investigation is aimed at understanding spheroid, and in particular bulge objects, the CM cut should bias heavily against the selection of disk stars.

K giants — the chosen spheroid tracer population — have absolute magnitudes in the range $-0.3 \lesssim M_V \lesssim 0.7$ and colours $1.0 \lesssim (B-V)_0 \lesssim 1.5$. So in the above CM selection window, the possible range of heliocentric distance these stars occupy is $5.7 \lesssim r_\odot \lesssim 23\,\mathrm{kpc}$. Along the chosen lines-of-sight, this heliocentric distance range spans from at least 1.7 bulge scale lengths in front of the line-of-sight tangent-point to about 30 scale lengths behind it. Integrating an $(r/0.9)^{-3.7}$ density law (*cf* Paper II) in a narrow cone along any of these lines-of-sight, one finds that this distance range contains at least 90% of all bulge stars in the cone. However, integrating an old disk double exponential $\exp(-R/3.5 - z/0.25)$ density law (see *e.g.* Gilmore, King & van der Kruit 1990 for an outline of our Galaxy model) over the same distance range, one obtains less than $\approx 20\%$ of the total old disk stars available along the line of sight. Thus the selection against K giants in the region along the line of sight from the Sun to 1.7 scale lengths in front of the line-of-sight tangent-point, loses at most 10% of bulge K giants, whereas it removes more than 80% of disk K giants. Furthermore, the vertical scale height of the disk ($\approx 250\,\mathrm{pc}$) is smaller than the vertical scale length of the 'outer bulge' (*cf* Paper II) or that of the halo, so that beyond the line-of-sight tangent-point a progressively larger ratio of bulge to disk stars will be found (except, of course, if the line-of-sight points along the galactic plane). Hence spheroid stars will be selected out to large heliocentric distances.

A consequence of this selection function is that the intrinsically most luminous K giants will be detected preferentially behind the galactic centre. We allow for this bias, which turns out not to be important, in the analysis using star count models in the accompanying paper.

A potentially larger effect is that the number of stars with intrinsic colour bluer than our cutoff which are scattered into the sample by photometric errors will exceed the number with redder intrinsic colours which are scattered out. Again, this bias is accounted for in our modelling.

The reddest M stars are introduced into the spectroscopic sample for the reason stated in section 1.2. The M giants can be distinguished from other stars in the sample in the way described in section 8 (one cannot simply impose a colour cut $(B-V)_0 \gtrsim 1.5$, because the *rms* errors on $(B-V)_0 \approx 0.2$ — *cf* section 2.4) are too large. M giants have absolute magnitudes $-0.6 \lesssim M_V \lesssim -0.2$ and colours $1.5 \lesssim (B-V)_0 \lesssim 1.8$, so in the above CM selection window, they lie in the heliocentric distance range $8.7 \lesssim r_\odot \lesssim 26\,\mathrm{kpc}$. Therefore the proportion of M giants that are spheroid rather than disk objects is larger than in the K giants in the chosen regions. Although M dwarfs cannot be easily distinguished from M giants with the data obtained, the many relevant surveys establish the M dwarf surface density at these apparent magnitudes and colours to be at most 2 stars $\mathrm{deg}^{-2}$ in the chosen regions, so that less than 2% of stars beyond $(B-V)_0 = 1.5$ (and with $14.5 \lesssim V \lesssim 16.5$) are expected to be dwarfs. It is concluded that with the above CM selection and in the regions studied, foreground disk contamination is not significant, so that the whole M-star sample kinematics trace spheroid kinematics to a good approximation.

The K giant sample is expected to be contaminated by disk dwarfs ranging from reddened early–K dwarfs ($(B-V)_0 \approx 0.8$, $M_V \approx 6$) to mid M dwarfs ($(B-V)_0 \approx 1.5$, $M_V \approx 10$). These stars account for at most $\approx 20\%$ of the selected K stars according to our Galaxy model. 21% of the spectroscopic sample are classified as dwarfs in section 8, so the estimates are consistent with each other and lend sup-port to the Galaxy model. The largest heliocentric distance at which the dwarfs can still be picked up with a faint magnitude cutoff of V < 16.5 is 1.3 kpc. Thus the dwarfs are all foreground objects and should have a kinematic signature distinctive of the local disk. We use this in our modelling later.

### 1.4 Selection of Observed Wavelength Range

The Mg'b' feature at $\approx 5170\text{Å}$ has the greatest, understood, sensitivity to temperature, metallicity and surface gravity in K star spectra. Several workers (e.g. Faber et al.1985, Friel 1986) have defined spectroscopic indices near the Mg'b' feature and calibrated them on a large number of standard stars. To make use of this information, spectra taken for this project were observed between $4600 \lesssim \lambda \lesssim 5600\text{Å}$. The most distinctive absorption features in this range are detailed in section 8.2. Additionally, this wavelength range and resolution is almost identical to that used by Kuijken & Gilmore (1989), so that their spectroscopic standards can be used to supplement those observed for this study (*cf* section 8).

## 2 PHOTOMETRY

### 2.1 Photographic Photometry

The several southern sky $B_J$, R and (where available) I UK Schmidt Telescope (UKST) plates that cover the 18 regions of sky at the intersections of $\ell$=(-25°, -15°, -5°, 5°, 15°, 25°) and $b$=(-12°, -15°, -20°) were scanned on the Automated Plate Measuring machine (APM) at Cambridge.

The APM is a high speed microdensiometer designed to measure the density of exposed emulsion on a photographic plate. A detailed account of how the APM detects images on plates, finds their magnitudes and copes with the many related problems such as overlapping plates and image classification, is given by Maddox (1988), and need not be repeated here. The *rms* magnitude error of 5% quoted by Maddox (1988) for repeated scans of UK Schmidt plates translates into an error $\delta(B - V) \approx 0.1$ in the colour-magnitude selection region detailed in section 1.3. However, the *rms* colour errors can be lower over small regions of the plate.

### 2.2 Comparison of APM to CCD Photometry

Digitised Schmidt plates are an extremely efficient tool for obtaining positions and relative magnitudes of a large number of low surface density objects. However, the instrumental APM magnitudes must be calibrated onto a well defined photometric scale over the magnitude range of interest, and any zero point differences introduced by plate matching must be corrected for. Additionally, it is desirable to check the effect of the plate resolution on the derived photometry.

Problems with APM photometry stem primarily from the limited dynamic range of the machine and the non-uniform response of the plates. CCDs are better photometers due to their linear response, but cover a much smaller solid angle on the sky. For example using the wide-angle Anglo-Australian Telescope Thomson f/1 focus CCD camera, some 250 CCD frames would be needed to cover all the fields for which we obtained spectroscopic data.

In crowded fields CCDs also have the advantage of better resolution: Schmidt plate resolution is $\approx 3"$, as opposed to $\approx 1-1.5"$ with the CCD observations detailed in section 2.3. To show the importance of this last point, take our most crowded CCD frame (*cf* section 2.3), in which the average separation between detected images is $\approx 7"$. Upon simulating a 3" instrumental resolution, it is found that overlapping images cause an *rms* error of $\approx 0.06$ in (V-I) colours for stars with 14.5 < V < 16.5.

### 2.3 CCD Photometry Observations and Data Reduction

CCD photometry was obtained in V and I on the Anglo-Australian Telescope during September 24-25, 1990, by Rosemary Wyse and GG, using the Thomson f/1 CCD camera. This $1024^2$ pixel configuration gives wide-field (17 arcmin$^2$) images. Five CCD fields in three of the 18 APM photometry fields were observed: one pair of images was obtained for the $(-25°, -12°)$ region, three pairs for the $(5°, -12°)$ region and one pair for the $(5°, -15°)$ region. This was supplemented in some fields by a preliminary calibration obtained by J.R. Lewis on the Mount Stromlo 40 inch Telescope under non-photometric conditions. The reduction of the AAT data is detailed in the remainder of this section.

The following frames were exposed: five pairs of bulge fields in V and I (at the above mentioned positions), a dark current frame, several photometric sequences (detailed below), ten and six bias frames on the first and second night respectively and three sets of V and I sky flats. Exposures of 5 sec were needed for the bulge fields to obtain photometry from V=13.0 to V=21.1 and from I=12.3 to I=19.8.

The IRAF software packages for crowded field photometry, DAOPHOT and PHOTCAL, were used to reduce the images. For a detailed description see Massey & Davis (1992).

The CCD frames are first de-biased. Variations in the sensitivity over the chip are removed by dividing the data frames by flat field frames constructed from offset-sky frames. Three sets of offset-sky frames, composed of (8,7), (5,5) and (5,5) frames in (V,I), were observed. Each of these sets is combined and then normalised to give a flat-field frame in V and I. The data frames are divided by a normalised sky flat (made from sky frames taken closest in time to the data frame and in the same filter). The sky regions in the data frames are now flat to better than 1%.

Typically 10000-20000 stellar images were detected in the $1024^2$ pixel bulge frames; a large proportion of these images are merged. In each frame, 10-20 apparently uncrowded stars were chosen from locations spread as uniformly as possible over the frame. These stars sample the point spread function (psf) of the stellar images and the average of the corresponding point spread functions defines the composite-psf. At each of the detected stellar image locations in the frame, the composite psf is scaled to the stellar intensity, and subtracted from the frame. If good psf-sampling stars have been chosen and the psf is spatially invariant, the resulting frame should contain only random noise and images of any previously undetected (and usually merged) stars. Any previously undetected images are now measured and subtracted from the frame, as before. If the resulting 'star-subtracted' frame has a variation comparable to the variation of areas of background sky in the original frame, the composite psf is accepted, otherwise a new set of uncrowded stars is chosen and the process is repeated. The accepted psf is scaled to give a fit to the stellar images, from which instrumental apparent magnitudes are calculated.

Aperture photometry is performed with the program PHOT on photometric sequences selected from Stobie, Gilmore & Reid (1985) and Stobie, sagar & Gilmore (1985), Christian et al.(1985) and Hawkins & Bessel (1988). The relations between instrumental magnitudes mV, mI, airmass XV (in filter V), XI (in filter I), colour (V − I) and magnitudes V, I were found, by least squares fitting, to be

$$\text{mV} = \text{V} + (0.555 \pm 0.013) + 0.170\text{XV}$$
$$+ (-0.0344 \pm 0.014)(\text{V} - \text{I}), \quad (1a)$$

and

$$\text{mI} = \text{I} + (1.169 \pm 0.013) + 0.080\text{XI}$$
$$+ (-6.910^{-4} \pm 0.011)(\text{V} - \text{I}), \quad (1b)$$

with $rms$ residuals of 0.03 and 0.04 mag respectively. The airmass - magnitude slopes of 0.170 in V and 0.080 in I were adopted from previous experience, as too few observations at high airmass were available for a direct determination.

A aperture correction is then applied to the psf photometry to reproduce the same zero point as the photometric sequence aperture photometry. Finally, bulge field V and I magnitudes are calculated by inverting equations 1a and 1b.

### 2.4  Calibration of APM Photometry

Colour equations are now needed to relate the CCD V and I magnitudes to the APM $B_J$ and R. The necessary relations are derived by fitting fifth order polynomials to photometry from Landolt (1983) (for $(B - V)_0 < 1.0$) and photometry of K giants selected from Menzies et al.(1980,1989) (for $(B - V)_0 \geq 1.0$); B is converted into $B_J$ using $B - B_J = 0.28(B - V)$ (Blair & Gilmore 1982). The several colour equations that will be needed for future use are displayed, with polynomial fits and corresponding $rms$ errors, in Figure 11 in appendix A.

This CCD derived $B_J$ and R photometry is used to calibrate the photographic photometry described in section 2.1. Stars common to both data sets are displayed in Figure 1; the straight line $\text{mag}_{CCD} = \text{mag}_{APM}$ is drawn to aid the eye, and the adopted fifth order polynomial fit is superimposed. The $rms$ errors in the $(B_J, R)$ fits are $(0.24, 0.09)$, $(0.25, 0.11)$ and $(0.19, 0.16)$ in the $(\ell = -25°, b = -12°)$, $(\ell = 5°, b = -12°)$ and $(\ell = 5°, b = -15°)$ regions respectively. For comparison to the galaxy model, a conversion is also needed from the APM $B_J$ and R into Cousins B and V: the necessary colour equations are derived in the same way as those in Figure 11 in Appendix A.

Finally, given the $rms$ errors in the CCD photometry, the $rms$ errors in the calibration of the APM photometry and the $rms$ errors in the colour equations, we find that the external $rms$ errors of the APM photometry when converted into $(B - V)$ are 0.18, 0.19 and 0.18 magnitudes in the $(\ell = -25°, b = -12°)$, $(\ell = 5°, b = -12°)$ and $(\ell = 5°, b = -15°)$ regions respectively.

We also obtain a very approximate calibration for the $(\ell = -5°, b = -12°)$ field by requiring that the position of the giant branch in that field lies at the same $(B - V)_0$ as that in the field at $(\ell = 5°, b = -12°)$. Mapping the ridge of the giant branch in narrow intervals of apparent V magnitude shows that any deviation is less than $(B - V) = 0.1$, that is, it is smaller than the $rms$ error in the photometry.

## 3  SPECTROSCOPIC DATA AND REDUCTION

Most of the spectroscopic data for this project were gathered during three runs at the Anglo-Australian Observatory in New South Wales, Australia on the 3.9m Anglo-Australian Telescope (AAT). During two observing runs from 20-25 June 1989 and from 9-13 July 1990, the optical fibres system AUTOFIB, the RGO spectrograph and the Image Photon Counting System (IPCS) were used. The more efficient TEK $1024^2$ CCD replaced the IPCS detector during an observing run from 3-8 August 1993. Additionally, we have used some 30 dwarf and giant spectroscopic standards which were observed earlier by Kuijken and Gilmore on the AAT multi-object FOCAP system with a similar setup, and reported in Kuijken & Gilmore 1989.

AUTOFIB offers a 40 arcmin diameter field on which one can position up to 64 optical fibres for simultaneous exposure. All regions described in §1.1 have a high enough K giant surface density in the CM selection window that at least 450 candidate K giants are available per fibres field. On a typical survey-region exposure, 40 - 50 candidate stars were observed, the remaining fibres (except for 4 broken ones needed for reliable sky subtraction) were allocated to regions of 'background sky'. The spectra were sampled at 1024 pixel locations between $4600 \lesssim \lambda \lesssim 5600 \text{Å}$ (*cf* section 1.4); integrations of approximately 4000-6000sec with the IPCS setup and $\approx$ 2000-3000 sec with the CCD setup are required to achieve signal to noise $\gtrsim 10$. (The actual wavelength ranges used were $4609 \lesssim \lambda \lesssim 5659 \text{Å}$, $4637 \lesssim \lambda \lesssim 5606 \text{Å}$, $4779 \lesssim \lambda \lesssim 5590 \text{Å}$ in the June 1989, July 1990 and August 1993 runs respectively).

A total of 40 survey fields was observed, including fields repeated from one observing run to the next. After each survey frame, three 'background sky' frames were observed with the same fibres configuration as used for the survey frame, but with the field-centre displaced by 10 arcsec along a cardinal direction. These background sky frames are used in the sky-subtraction algorithm detailed in section 3.2. To obtain survey star radial velocities by crosscorrelation with a template (section 4), 24 radial velocity standard stars (whose velocities were taken from the Astronomical Almanac) were observed. Twilight sky frames were exposed at the beginning and end of each night to give an external measure of velocity errors (section 5). To calibrate survey star metallicities (section 6), 14 K giant metallicity standards were observed. A K dwarf, a K giant and an M dwarf spectral standard were observed to help classify the survey stars (section 8). Arc-lamp frames were observed both immediately before and after each survey frame and either immediately before or after observing a radial velocity standard star or the twilight sky (section 3.1).

### 3.1  Reduction

The CCD frames are de-biased, and $5\sigma$ cosmic ray events more than 50 counts over the local background are removed using the FIGARO routine BCLEAN. The 64 fibres spectra lie in curves, typically 4-5 pixels wide, over the IPCS and CCD frames. The ridge of each of the 64 curves is found by observing a tungsten lamp (a uniform source), and polynomial fits are made along the ridges. Spectra are extracted by summing up counts in a band centred on the polynomial fit and locally normal to it, with the spectra extracted using optimal (local signal-noise ratio weights applied) weighting. The adopted width of the extracted region was 4.75 pixels for all frames.

**Figure 1.** The CCD photometric calibrations. All stars in common to the CCD and APM photometry lists are plotted. V and I CCD magnitudes have been converted into APM instrumental $B_J$ and R as detailed in the text. A $2\sigma$ clipped fifth order polynomial fit is superimposed. (a), (b) are for the region ($\ell = -25°, b = -15°$), (c), (d) are for ($\ell = 5°, b = -12°$) and (e), (f) are for ($\ell = 5°, b = -15°$).

The data frames are binned logarithmically to obtain radial velocities (since a velocity shift between two spectra is proportional to $\Delta\lambda/\lambda = \Delta\ln\lambda$) and linearly to obtain metallicity estimates (since spectral indices have been defined on linear data). To find the transformation from $x$-position along the spectra into wavelength, the survey frame exposures were interlaced with Cu-Ar and Fe-Ar arc lamp exposures. A seventh order polynomial fit is made between wavelength and $x$-position. The $rms$ error in these fits was always better than 0.14 Å (or $\approx 8\,\mathrm{km\,s^{-1}}$); an average value for the fits of the 1993 data-set is 0.07 Å $rms$ ($\approx 4\,\mathrm{km\,s^{-1}}$).

### 3.2 Sky-subtraction

For a 3000 sec survey star exposure, a representative count per bin in fibres assigned to 'background sky' is ≈ 100, while survey star fibres exceed the background sky by ≈ 50 to ≈ 1000 counts per bin (the average number of counts per bin in the 1989, 1990 and 1993 data-sets is ≈ 150, ≈ 150 and ≈ 500 respectively). It is critical that survey star spectra, especially those of low S/N, have the contaminating sky spectrum removed for the following reasons. In velocity estimation, sky-lines in the survey star will correlate against sky-lines in the radial velocity standard star which could result in a mistaken radial velocity. In metallicity estimation, the largest effect with low signal to noise spectra is that both the stellar absorption features and the continuum are displaced by the mean sky count, thus reducing the relative depth of absorption lines from the continuum so that stars appear more metal-poor than they should be. Furthermore, sky emission lines can be as prominent as stellar absorption lines in very metal poor stars. This last consideration requires that the actual correctly-normalised sky-spectrum be removed, not just a mean sky-count.

The background sky spectrum was sampled during every survey star exposure (typically with 5-10 fibres), as the spectrum of the background sky varies with position on the sky and depends on many other variables such as zenith distance and the seeing conditions.

The sky-subtraction is performed with a FIGARO routine developed for this project and based on a recipe detailed in Wyse & Gilmore (1992). This also corrects for the spectrograph dark count, for light scattered uniformly over all fibres and for light scattered from one fibre into neighbouring fibres. The method is able to subtract sky emission to ≈ 2%, judging by the relative amount of flux in the 5577Å sky emission line compared to nearby continuum before and after sky-subtraction. Figure 2 shows a typical medium signal to noise spectrum of a candidate bulge object before and after sky-subtraction.

### 4 RADIAL VELOCITIES FROM CROSSCORRELATION

By applying standard crosscorrelation techniques (Tonry & Davis 1979) to these data, we obtain relative shift accuracies of $\sigma \approx 1/6$ of a pixel (*cf* section 5).

Before crosscorrelating the logarithmically binned spectra, a low order polynomial is fit and subtracted from the spectra, and a cosine bell applied to taper 5 percent at the beginning and end of the wavelength range. The relative shift calculated in the crosscorrelation between the sky-subtracted survey star spectrum or twilight sky spectrum with the velocity standard star spectrum is interpreted as a velocity difference $v = c\Delta \ln \lambda$. A heliocentric velocity correction is applied to both the survey stars and the radial velocity standard stars. The survey star heliocentric velocity $v_{\mathrm{HEL}}$ is then transformed into the Local Standard of Rest (LSR) velocity $v_{\mathrm{LSR}}$ by:

$$v_{\mathrm{LSR}} = v_{\mathrm{HEL}} + V_\odot (\cos\alpha_\odot \cos\delta_\odot \cos\alpha \cos\delta$$
$$+ \sin\alpha_\odot \cos\delta_\odot \sin\alpha \cos\delta$$
$$+ \sin\delta_\odot \sin\delta)\,\mathrm{km\,s^{-1}}, \quad (2)$$

**Figure 2.** A typical medium signal to noise ratio spectrum before (top curve) and after sky-subtraction. A bias of 100 counts has been added to the top curve so as to avoid overlap. The success of the sky-subtraction in this example can be seen in the clean removal of the sky emission line at 5577Å.

and from the LSR frame to a Galactic frame (in which the LSR rotates at the circular velocity of the Galactic potential at the solar neighbourhood) by:

$$v_{\mathrm{GAL}} = v_{\mathrm{LSR}} + v_0 \sin\ell \cos b, \quad (3)$$

where $\alpha$, $\delta$ and $\ell$, $b$ are the equatorial and galactic coordinates of the survey star, $v_0 = 220\,\mathrm{km\,s^{-1}}$ is the circular velocity at the solar radius and $V_\odot = 16.5\,\mathrm{km\,s^{-1}}$ is the solar motion in the direction $(\alpha_\odot, \delta_\odot) = (18.0°, 30°)$ relative to stars in the solar neighbourhood. Henceforth, we shall denote radial velocities in this Galactic frame as 'Galactocentric radial velocities'.

### 5 VELOCITY MEASUREMENT ERRORS

Many factors such as variations in seeing conditions, sky intensity and airmass will give rise to errors in the velocity measurements. Instrument dependent factors such as the change of detectors (section 3), the slight changes in the observed wavelength range (detailed in section 3) and errors in arc lamp calibrations will also alter the derived kinematics. The effects of all these variables are difficult to quantify directly, so instead we estimate these measuring errors from repeated observations of the same objects during different observing runs. Figure 3 shows the distribution of the absolute values of the differences in velocity between all pairs of repeated observations; superimposed is a maximum-likelihood fitted gaussian that has $\sigma = 8.4\,\mathrm{km\,s^{-1}}$ (the KS test probability for this fit is 0.24, so the fit is good). Since the value of $\sigma$ calculated from Figure 3 is the standard deviation of the measurement $d = |(v_1 - v_2)/2|$, then $\sigma_v = \sqrt{2}\sigma_d$, so that $\sigma_v = 11.9\,\mathrm{km\,s^{-1}}$. Thus arc-lamp calibration errors do not account for the bulk of the velocity error (the mean $rms$ wavelength calibration error is ≈ 4 km s$^{-1}$, from section 3.1); the remainder arises from limited S/N, with some contribution from star-template mismatch.

Twilight sky observations are frames in which all but the broken fibres record the spectrum of scattered sunlight. The comparison of velocities from twilight sky frames taken at intervals during the course of an observing run provides a way to check the temporal stability of the velocity zero points; for instance, it is desirable to know what velocity errors should be expected from the crosscorrelation of spectra taken on different nights or on different observing runs. To do this, the highest S/N twilight sky spectrum ($S/N \approx 14$) taken on the June 1989 observing run is crosscorrelated against the other spectra from the same frame, giving a mean velocity difference of $4.0\,\mathrm{km\,s^{-1}}$ with an $rms$ error of $10.6\,\mathrm{km\,s^{-1}}$. This result does not vary significantly even when crosscorrelating against spectra taken on different observing runs: the above spectrum crosscorrelated against the total of 12 twilight sky frames taken during all three observing runs gives a mean velocity difference of $6.5\,\mathrm{km\,s^{-1}}$ with an $rms$ error of $11.2\,\mathrm{km\,s^{-1}}$.

**Figure 3.** The distribution of the absolute values of the differences in velocity from the mean between all pairs of repeated observations; superimposed is a maximum-likelihood fitted gaussian that has $\sigma = 8.4\,\mathrm{km\,s^{-1}}$.

It is possible that damaged fibres, or the difference in response over the fibres frame (section 3.2), could cause systematic variations in measurement errors across the fibres frames. This is investigated by finding, at each fibre position, the average and $rms$ error of the velocities from the 12 twilight sky frames discussed above (the velocities are found by crosscorrelation against the highest S/N twilight sky spectrum observed in the June 1989 run). The result is displayed in Figure 4; the line $v = 6.5\,\mathrm{km\,s^{-1}}$ (the mean velocity of the template, see previous paragraph) and the broken fibres (marked 'dead') are superimposed on the plot. The figure shows that measuring errors are consistent with being uniform over the fibres frames.

External errors are estimated by cross-correlating a radial velocity standard star against all the twilight sky spectra. The resulting velocities are corrected for the heliocentric velocity and the radial velocity of the standard. Thus the velocity of the twilight sky is $-3.0 \pm 4.1\,\mathrm{km\,s^{-1}}$, which is consistent with being zero.

## 6 METALLICITY ESTIMATES

K stars have spectra that are primarily dependent on temperature (colour), surface gravity and metallicity. However, since giants can be reliably discriminated from dwarfs (*cf* §8.2), most of the surface gravity dependence in the K giant spectra can be removed. Then, given the colour of a star, an abundance estimate can be obtained by comparing one or more line strength indices against the same line strength indices obtained from metallicity standards of the same colour. Line-indices for $\approx 60$ metallicity standards were taken from Friel (1986) and Faber et al.(1985).

**Figure 4.** We show the mean and $rms$ velocity calculated from cross-correlating at each fibre position all twilight sky frames with the highest S/N twilight spectrum taken during the June 1989 observing run. The mean velocity of this template spectrum is superimposed as a broken line. The measuring errors are therefore consistent with being uniform over the fibres frames.

However, spectra taken with fibres systems are different from fluxed spectra (on which the line strength indices are defined): repeat observations of any star give spectra with significantly different shapes. In the data available here the variations have a characteristic width larger than $\approx 200\mathring{A}$, and so cannot be confused with absorption features. Given that the shape of a low order polynomial fit to the spectrum of a particular star would thus contain erroneous information, the spectrum should first be flattened by dividing by such a fit. However, it is important to retain information on the deep molecular and atomic absorption bands in the K star spectrum. hence we fit the polynomial through the several continuum regions quoted by Friel (1986); those in the region $4600 \lesssim \lambda \lesssim 5600\mathring{A}$ are presented in table 2.

For the resulting continuum - divided spectra, we define an index analogous to the metallicity sensitive Mg index defined by Friel (1986), but which has no continuum normalisation:

$$\mathrm{Mg_{Flat}} = -2.5\log \int_{5130}^{5200} \frac{F_\lambda}{70\mathring{A}} d\lambda, \qquad (4)$$

where $F_\lambda$ is the intensity of the continuum - divided spectrum at wavelength $\lambda$. The relation between our Mg index and Friel's original Mg index (which, to avoid confusion, we will denote $\mathrm{Mg_{Friel}}$) is plotted in Figure 5 for the 14 metallicity standards for which we have spectra. This relation is:

$$\mathrm{Mg_{Friel}} = 0.7097 \pm 0.016 + (1.2612 \pm 0.074)\mathrm{Mg_{Flat}}. \qquad (5)$$

To make use of the standards observed by Faber et al.(1985), their $\mathrm{Mg_1}$ index (which we will denote $\mathrm{Mg_{Faber}}$) is trans-

**Table 2:** The continuum regions in the region $4600 \lesssim \lambda \lesssim 5600 \text{Å}$ defined by Friel (1986) and adopted in this work.

| Lower $\lambda(\text{Å})$ | Upper $\lambda(\text{Å})$ | Lower $\lambda(\text{Å})$ | Upper $\lambda(\text{Å})$ |
|---|---|---|---|
| 4796.00 | 4841.00 | 4935.00 | 4975.00 |
| 5051.00 | 5096.00 | 5220.00 | 5250.00 |
| 5288.00 | 5367.00 | 5307.25 | 5317.25 |
| 5356.00 | 5365.00 | 5430.00 | 5460.00 |

formed into $\text{Mg}_{\text{Friel}}$ by linearly fitting data for stars in common to both lists:

$$\text{Mg}_{\text{Friel}} = 0.7494 \pm 0.0067 + (0.836 \pm 0.026)\text{Mg}_{\text{Faber}}. \quad (6)$$

In Figure 6 the metallicity values of metallicity standard stars from Friel (1986) and Faber et al.(1985) are plotted in the $\text{Mg}_{\text{Friel}}$, $(B-V)$ plane. This surface was fit by a ninth order polynomial:

$$\begin{aligned}
[Fe/H] = \quad & -68.74 & -19.71(B-V) \\
& +139.3Mg & +41.48(B-V)Mg \\
& +40.06(B-V)^2 & -86.49(B-V)^2Mg \\
& -67.48Mg^2 & -23.87(B-V)Mg^2 \\
& +46.15(B-V)^2Mg^2 & \quad (7)
\end{aligned}$$

We use the two dimensional Kolmogorov-Smirnov (KS) test to check the fit, and find that the probability that the KS statistic $D$ should be greater than the calculated value if the fit and the data are drawn from the same distribution is 0.42. The fit can therefore be considered to be acceptable.

The quantity $\text{Mg}_{\text{Flat}}$ is calculated for all those stars in our survey which are classified as K giants, and transformed into $\text{Mg}_{\text{Friel}}$ using equation 5. From $\text{Mg}_{\text{Friel}}$ and $(B-V)_0$ (from APM photometry and the reddening, *cf* §2), an estimate of the abundance is calculated via equation 7. Note that in Figure 6, the standard stars cover the survey star region adequately well. Note however that survey stars with interpolated metallicity $[\text{Fe/H}] \gtrsim 0.25$, or $[\text{Fe/H}] \lesssim -2.75$, or with $(B-V)_0 \gtrsim 1.5$ will not have a reliable abundance measure. It is also important to remember that our calibration of abundances onto an $[\text{Fe/H}]$ scale from observations of standard stars will be strictly appropriate to bulge stars only if the $[\text{Mg/Fe}]$ *vs* $[\text{Fe/H}]$ of the bulge is similar to that of the solar neighbourhood field.

## 7  METALLICITY MEASUREMENT ERRORS

### 7.1  Internal Errors

Internal metallicity errors are estimated from the absolute value of the difference in abundance of multiply observed stars. A histogram of these is shown in Figure 7; superimposed is a maximum-likelihood gaussian fit that has $\sigma = 0.26$, the KS test probability for this fit is 0.08, so the fit is fair but not excellent (see §6 for the interpretation of KS test probabilities), the actual distribution is more prominently 'winged' than a gaussian. Again, since the fit is performed on $d = |([\text{Fe/H}]_1 - [\text{Fe/H}]_2)/2|$, then $\sigma_{[Fe/H]} = \sqrt{2}\sigma_d = 0.37 \text{dex}$.

**Figure 5.** The relation between the metallicity-dependent index $\text{Mg}_{\text{Flat}}$ defined in the text and the Mg index $\text{Mg}_{\text{Friel}}$ defined by Friel (1986).

### 7.2  External Errors

External metallicity errors are estimated by modeling the effect of photon noise and colour errors on the standard metallicity star spectra.

Fourteen new spectra are constructed by degrading the S/N to $\approx 10$ — representative of the sky-subtracted survey spectra — of the 14 metallicity standard stars observed for this investigation. Photon noise is modeled using a Poisson random number generator (Press et al 1986). The spectra are continuum - divided to find the Mg index $\text{Mg}_{\text{Flat}}$, which is then converted into $\text{Mg}_{\text{Friel}}$ (these quantities are defined in §6); the error in $\text{Mg}_{\text{Friel}}$ is calculated from equation 5.

These 14 spectra are placed onto the $\text{Mg}_{\text{Friel}}$, $(B-V)_0$ plane of Figure 6. Assuming the *rms* colour error

$$\delta(B-V) = \sigma_{(B-V)} = 0.18 \quad (\S 2.4)$$

and that

$$\delta\text{Mg}_{\text{Friel}} = \sigma_{\text{Mg}_{\text{Friel}}}$$

, then the mean metallicity $\overline{[\text{Fe/H}]}$ and the standard deviation about this mean are calculated from

$$\overline{\left[\frac{Fe}{H}\right]} = \int_{-\infty}^{+\infty}\int_{-\infty}^{+\infty} P \; F \; d(B-V)_0 \; d\text{Mg}_{\text{Friel}}, \quad (8a)$$

$$\sigma^2_{[Fe/H]} = \int_{-\infty}^{+\infty}\int_{-\infty}^{+\infty} P \left(F - \overline{\left[\frac{Fe}{H}\right]}\right)^2 d(B-V)_0 \; d\text{Mg}_{\text{Friel}}, (8b)$$

where $P$ is a two dimensional gaussian probability function centred at

$$\left((B-V)_0, \text{Mg}_{\text{Friel}}\right)$$

with dispersions

$$\left(\sigma_{(B-V)}, \sigma_{\text{Mg}_{\text{Friel}}}\right)$$

and $F$ is the surface fit for $[\text{Fe/H}]$ in terms of

$$\left((B-V)_0, \text{Mg}_{\text{Friel}}\right)$$

**Figure 6.** The Index-colour plane for metallicity standards. Data are taken from Friel (1986) and Faber et al.(1985). The metallicity of the standard stars is represented by a symbol as described in the legend. Contours of equal metallicity are labeled in steps of 0.25 dex. The polynomial fit shown is detailed in the text. The 'small dots' show the $(B - V)_0$, $Mg_{Friel}$ positions of the stars in the present survey which are classified as K giants. The standard stars cover most of the $(B - V)_0$, $Mg_{Friel}$ region covered by the survey stars.

**Figure 7.** The distribution of the absolute value of the difference between independent estimates of the abundance from the mean of multiply observed stars. Superimposed is a maximum-likelihood gaussian fit that has $\sigma = 0.26$dex.

(equation 7).

Figure 8 shows the variation in the calculated error $\sigma_{[Fe/H]}$ as a function of [Fe/H]. The error on stars on the low metallicity end is higher because [Fe/H] changes rapidly with absorption line strength at low metallicities (the contour lines in Figure 6 are more closely stacked at the low metallicity end). The external errors from Figure 8 are consistent with the value obtained above for the internal error $\sigma_{[Fe/H]} = 0.37$dex.

Figure 9 shows the metallicity of the original standard plotted against the calculated $\overline{[Fe/H]}$; the line

$[Fe/H]_{Calculated} = [Fe/H]_{Standard}$

is superimposed. Using $\sigma_{[Fe/H]}$ from Figure 8, $\chi^2$ is calculated from

$$\chi^2 = \sum \left( \frac{[Fe/H]_{Standard} - [Fe/H]_{Calculated}}{\sigma_{[Fe/H]}} \right)^2. \quad (9)$$

The value of $\chi^2$ is thus 28.7; the probability that the fit should be as poor as this by chance (calculated from the incomplete gamma function) is 0.004. However, if the metallicity standard at $[Fe/H]_{Standard} = -2.5$ is removed — this metallicity standard is one of the rare metal poor giants of red colour $((B - V) = 1.2)$ — then $\chi^2$ becomes 17.5 and the probability for the fit is 0.1, which is (just) acceptable.

In summary, above $[Fe/H] = -1$, the metallicity errors are less than approximately 0.5 dex, but below this value we have too few standard stars to allow a check on the calibration.

**Figure 8.** The $\sigma$ error in [Fe/H] as a function of [Fe/H], estimated by calculating [Fe/H] (as prescribed in §6) for metallicity standard stars with modeled photon noise and colour errors.

**Figure 9.** The relation between the real metallicity of the standard stars and a mean metallicity calculated by modeling photon noise and colour errors.

## 8 SPECTRAL CLASSIFICATION

### 8.1 Recognising Carbon Stars and M giants

In the APM colour-magnitude window imposed on the stellar selection we expect to find K stars, some late-M stars and Carbon stars. A brief summary follows of how carbon stars and M giants are distinguished 'by eye' from one another and from the rest of the sample.

Carbon stars have striking spectra because the $C_2$ molecule produces absorption bands that are unusual in that the sharp edge of the band is on the red end of the absorption feature. All carbon star types displayed in Turnshek et al.(1985) strongly exhibit one or more of three $C_2$ absorption bands in the wavelength range 4600 to $5600\text{Å}$: the red edges of these bands are located at 4737, 5165 and 5636 Å. Most M giants are also easy to identify by eye, as they have deep TiO (and less prominent VO) absorption bands that drop very sharply on the blue end of the absorption band (*cf* e.g. Turnshek et al.(1985)). In the wavelength range 4600 to $5600\text{Å}$, TiO features occur at 4626, 4761, 4804, 4848, 5241, 5448, 5494, 5569 and 5598 Å. Using these criteria, carbon stars and M giants are identified and separated from the sample.

### 8.2 Surface Gravity Classification

The remaining stars in the sample are expected to be K stars and M dwarfs. K dwarfs can be separated crudely from M dwarfs by their colours, since M dwarfs have

$$1.4 \lesssim (B-V)_0 \lesssim 1.6$$

and K dwarfs have

$$0.8 \lesssim (B-V)_0 \lesssim 1.4$$

(Mihalas & Binney 1981), though recall the *rms* error in $(B-V) = 0.18$.

A grid of K giant, K dwarf and M dwarf standards covering a range of $(B-V)_0$, and taken from Kuijken & Gilmore (1989) is shown in Figure 10. The most striking features in these spectra are the three Mg'b' lines at (5167, 5173 and 5184 Å) and the MgH band at 5211 Å (which also belongs to the Mg'b' feature), the other lines being mostly TiO, FeI and FeII. Several properties of K star atmospheres can be seen in the grid. In dwarfs, a prominent MgH band is seen after $(B-V)_0 \gtrsim 1.05$, while in giants it appears only after $(B-V)_0 \gtrsim 1.25$. Also, even super-metal-rich giants have weaker Fe lines than Hyades dwarfs at the same colour.

Cayrel et al.(1991) calculate synthetic spectra to find the surface gravity dependence of a K star spectrum in the wavelength range 4800 to $5300\text{Å}$ at fixed effective temperature and metallicity. In this situation they show that dwarfs display much stronger Mg, Fe and MgH lines than giants, because giants have lower surface gravity atmospheres and hence lower opacities. Cayrel et al.also calculate the metallicity dependence at constant surface gravity and effective temperature — as would be expected, higher metallicity increases the depth of the Mg and Fe lines and the MgH band, except for saturated Mg lines in metal poor dwarfs. Their results show clearly that the Mg'b' triplet and MgH band are more sensitive to gravity than to metallicity for stars of $[Fe/H] \gtrsim -1.25$, and that these lines can be as weak in metal poor dwarfs as they are in giants. Fortunately, along the lines of sight we study, we expect from our Galaxy model to find a negligible number ($< 0.01$ %) of metal poor dwarfs (foreground halo stars) in the samples — see also §3.2.

Keeping the above remarks in mind, K giants, K dwarfs and M dwarfs are classified 'by eye' by comparison to the standards in Figure 10. The spectra were also binned into four groups, ranked in order of how sure we were of the classification. The giant-dwarf classification was deemed to be satisfactory for high S/N spectra, but was clearly unsatisfactory for noisy spectra, judging from our repeated attempts at classification, especially on the bluer end of the selection range ($(B-V)_0 \lesssim 1.0$).

Both the concern at the reliability of classifying low S/N data and the tedious nature of the above task prompted Ibata & Irwin (in preparation) to develop an algorithm to discriminate giant and dwarf spectra. The technique they use is a variant of principal component analysis (PCA), and will be described elsewhere.

Finally, we recall that the astrophysical analysis of these data are presented in an accompanying paper.

**Figure 10.** The grid of standard stars for luminosity classification.

### ACKNOWLEDGEMENT

We thank Rosemary Wyse for her collaboration in providing the AAT CCD photometric data essential for this project.

### APPENDIX A.: COLOUR EQUATIONS

The colour equations used to convert CCD $(V-I)$ to the UKST-APM $(B_J - R)$ are shown in Figure 11. In Figure 11, panel (a) shows the $(V-I)/(B_J-R)$ relation, while panel (b) shows the $(V-I)/(V-R)$ relation; the superimposed fifth order polynomial fits have an *rms* error of 0.033 and 0.009 magnitudes respectively. The fitting coefficients $a_i$ (defined by $y(x) = \sum_{i=0}^{5} a_i x^i$) are also displayed on the diagrams. The small dots are stars with $0.0 < (B-V)_0 < 1.0$ from Landolt (1983), and the large dots are E-Region K giants with $(B-V)_0 \geq 1.0$ taken from the catalogue by Menzies et al.(1980,1989).

### APPENDIX B.: THE DATA

The data are presented in a 62 page table, published as an accompanying microfiche. A single page is published here to illustrate the information available. The data are also available from the Centre de Donnees Stellaires, Strasbourg (http://cdsweb.u-strasbg.fr/CDS.htm1), and the Astronomical Data Center, NASA Goddard Space Flight Center (request@nssdca.gsfc.nasa.gov).

**Figure 11a.** Colour equations used throughout this study.

**Figure 11b.** Colour equations used throughout this study.

Wyse R., Gilmore G., 1992, MNRAS, 257, 1

This paper has been produced using the Blackwell Scientific
Publications TEX macros.

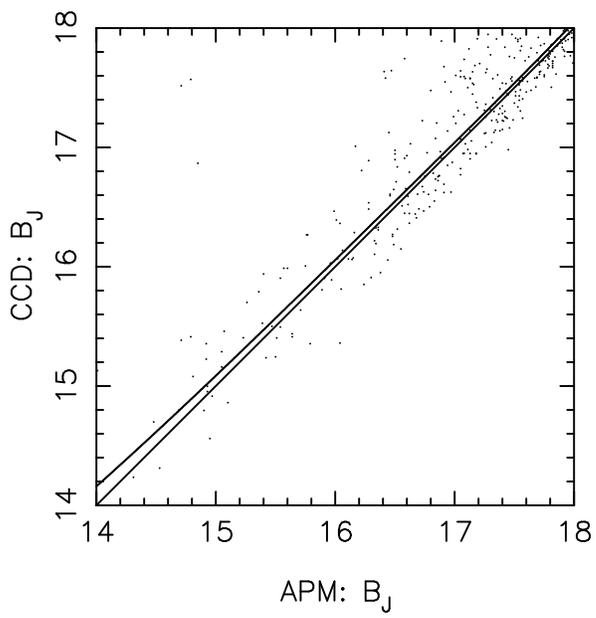
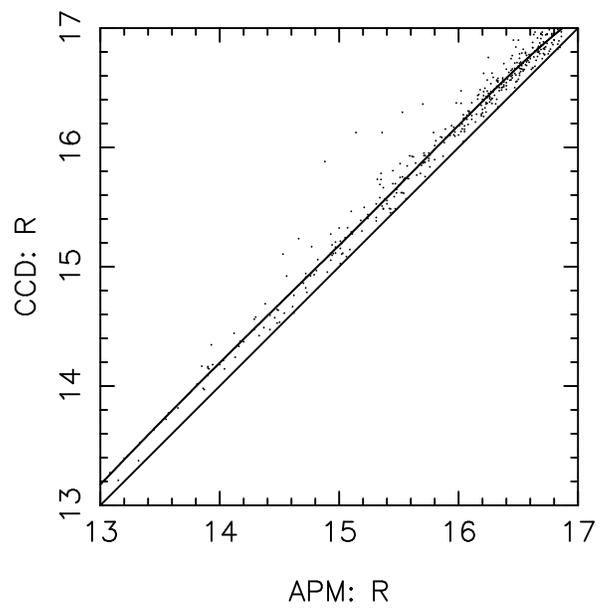

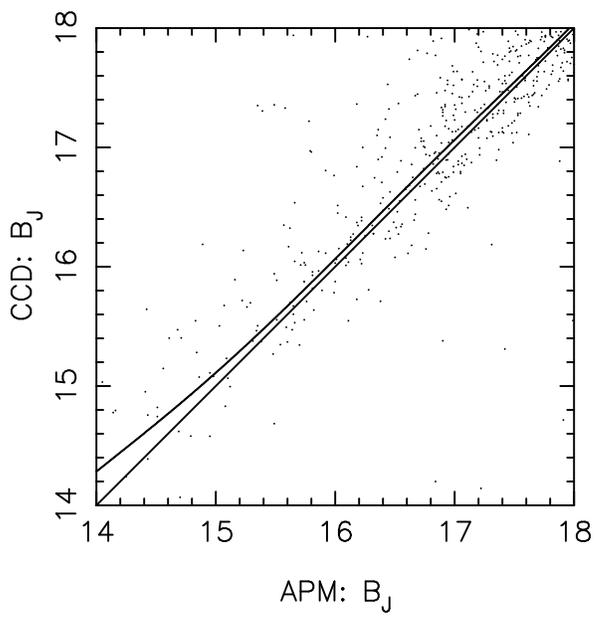 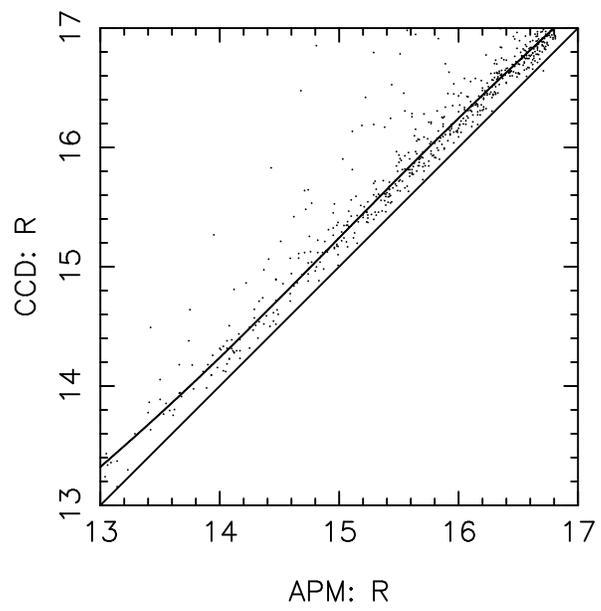

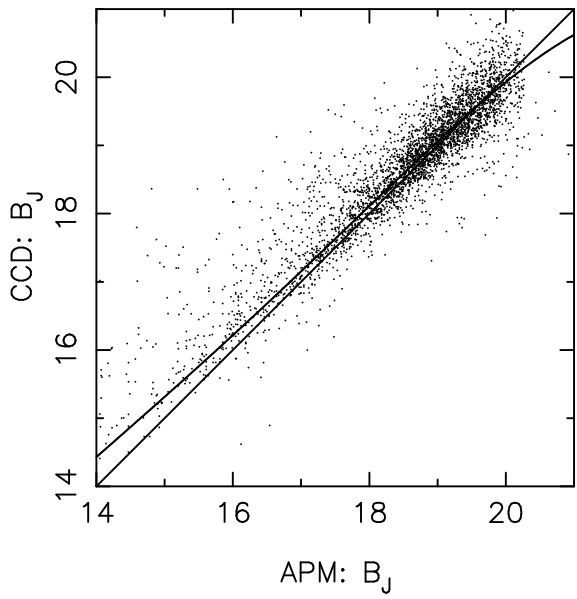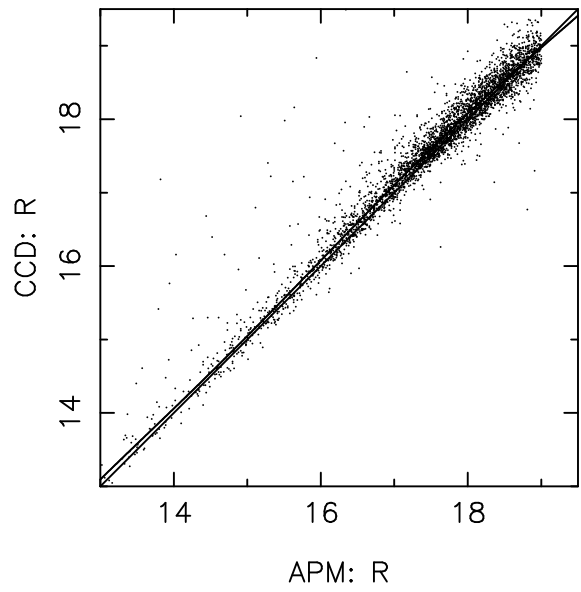

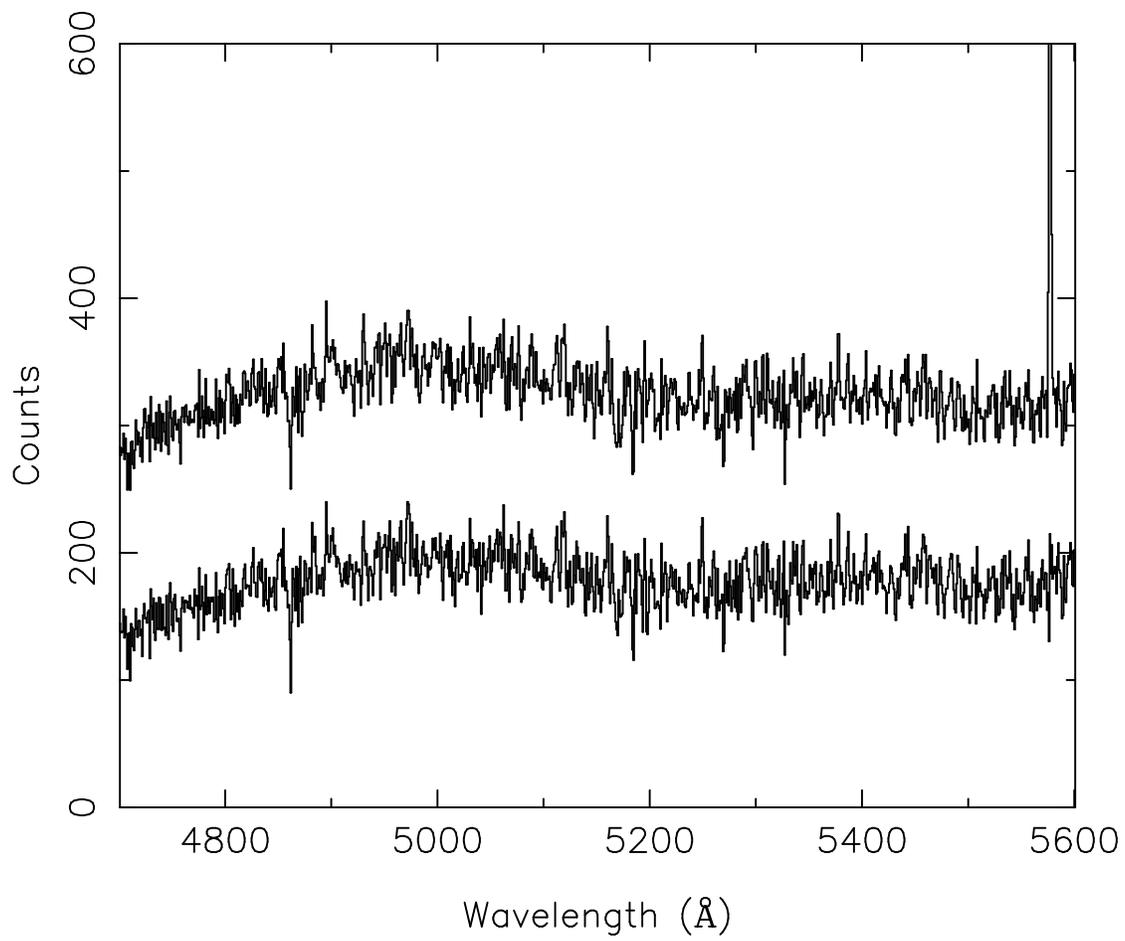

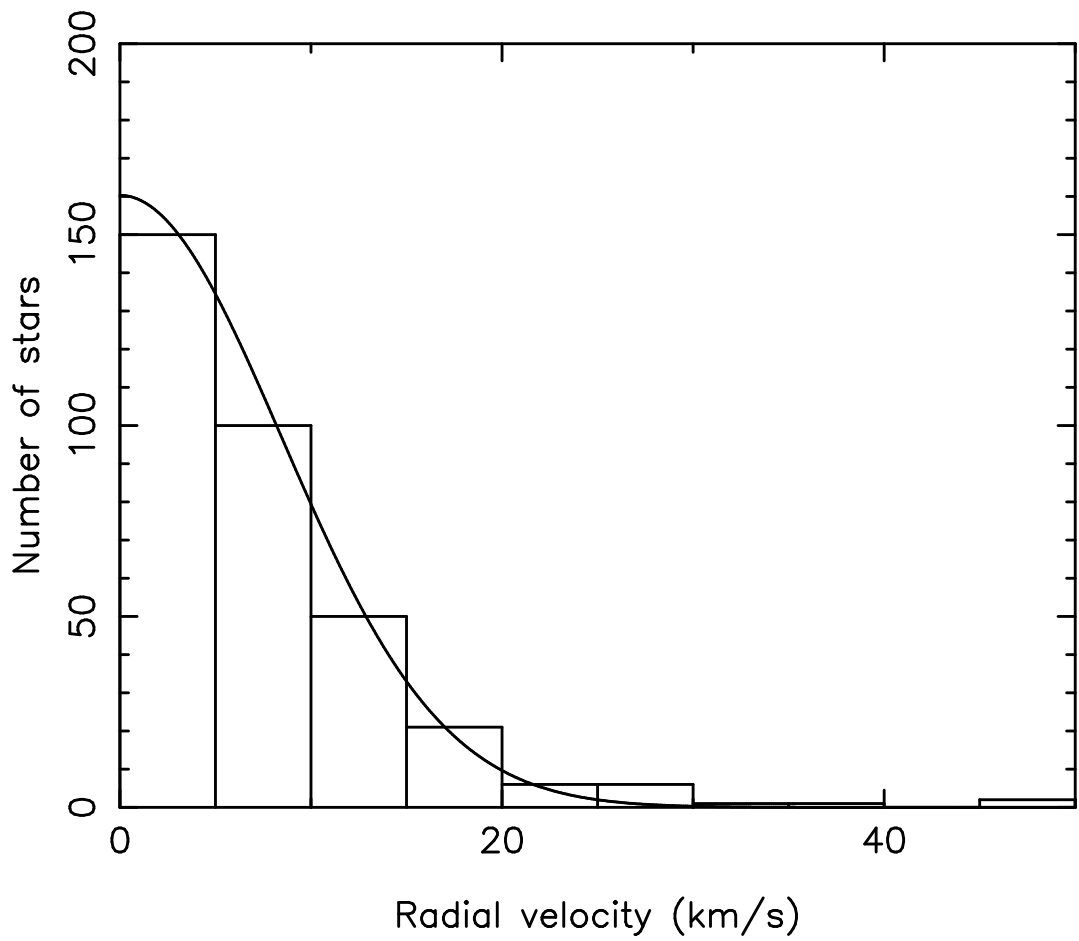

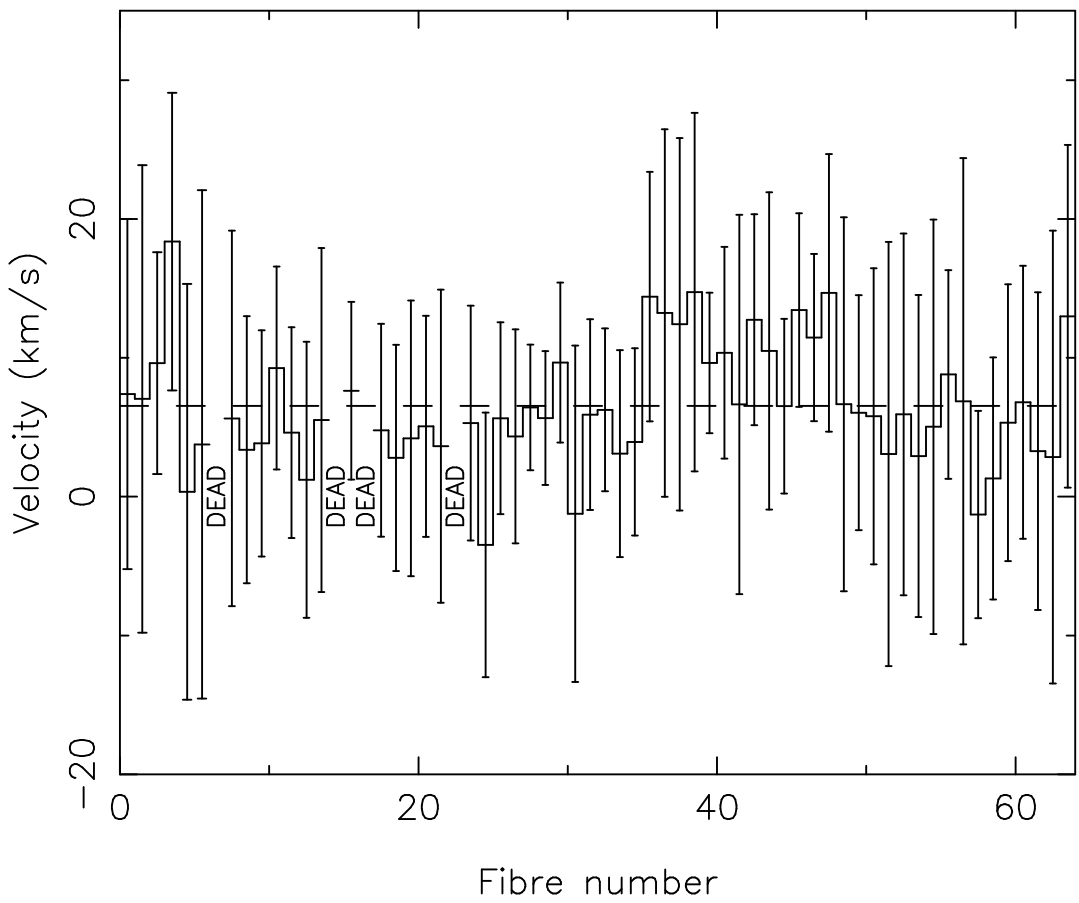

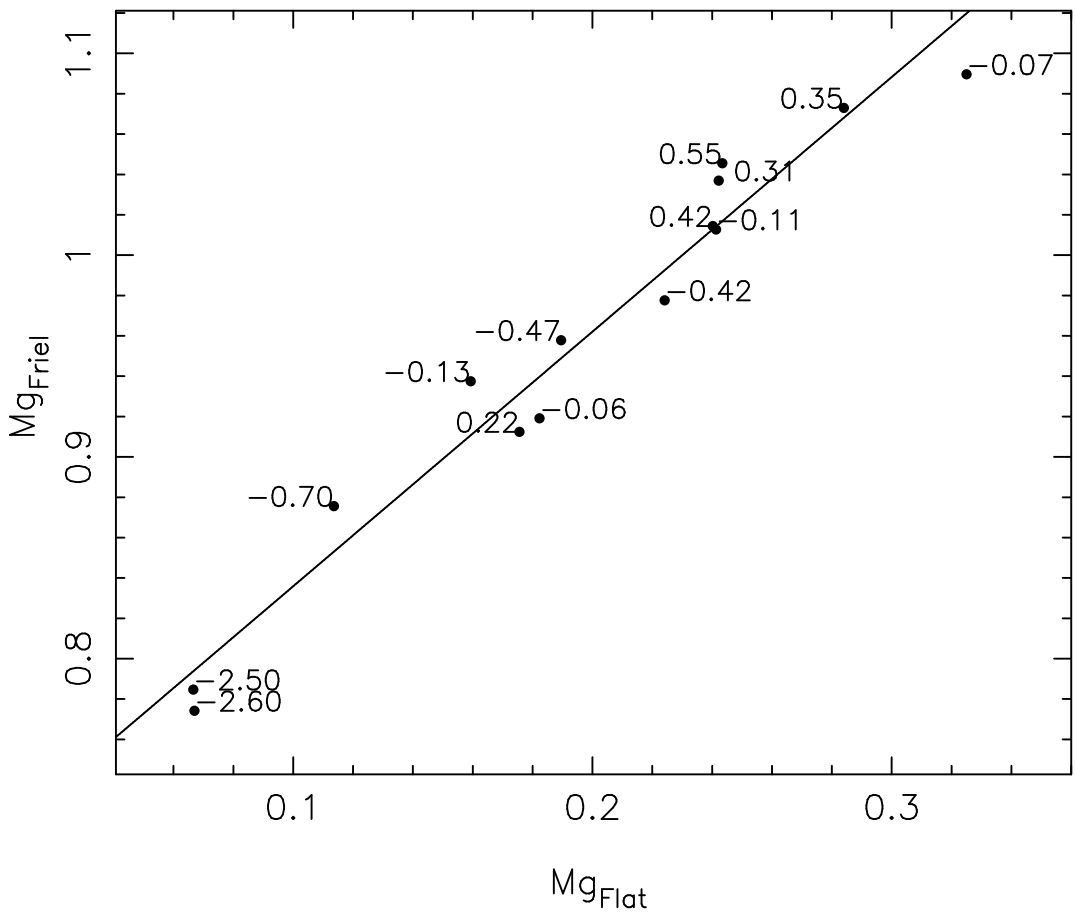

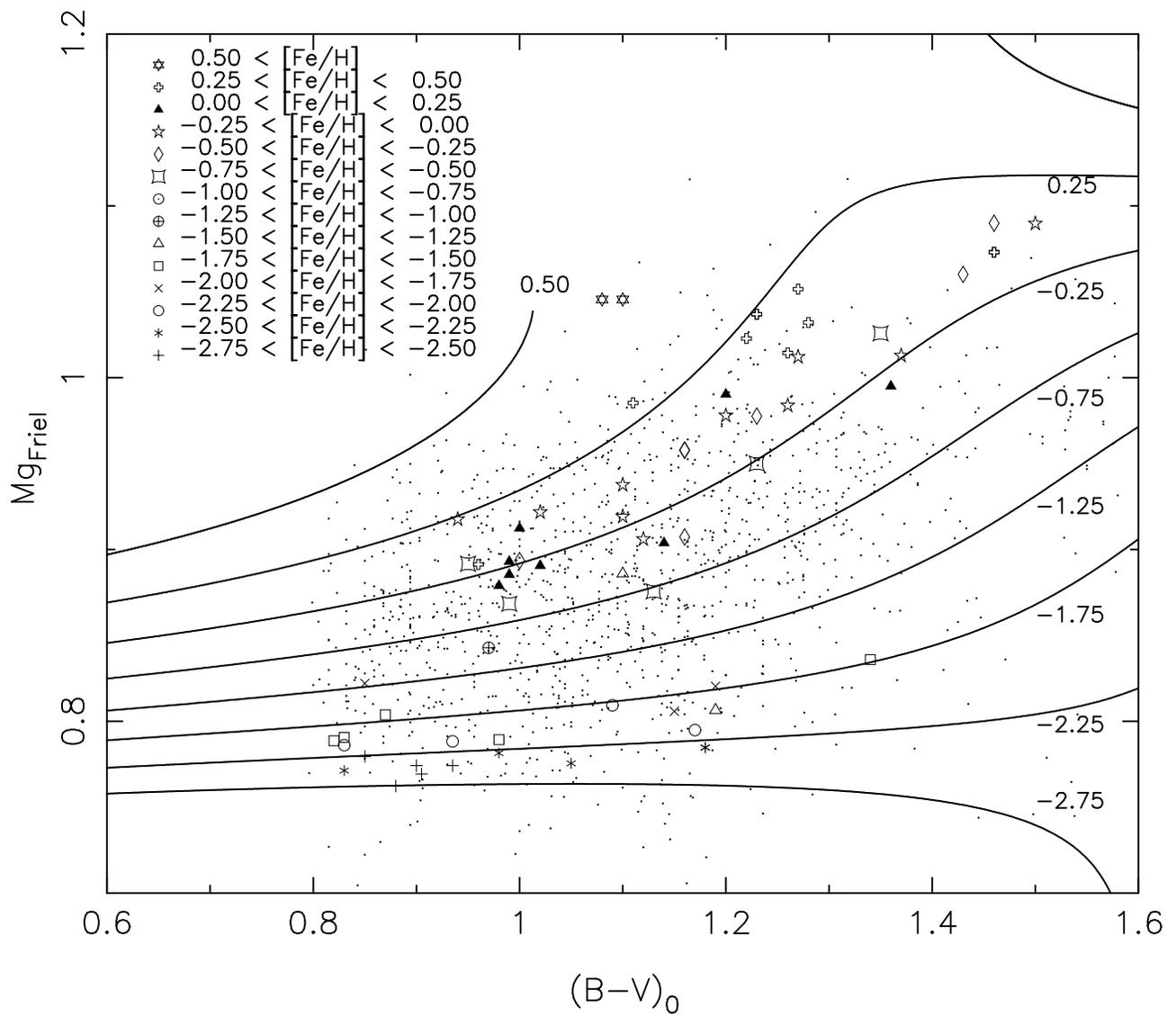

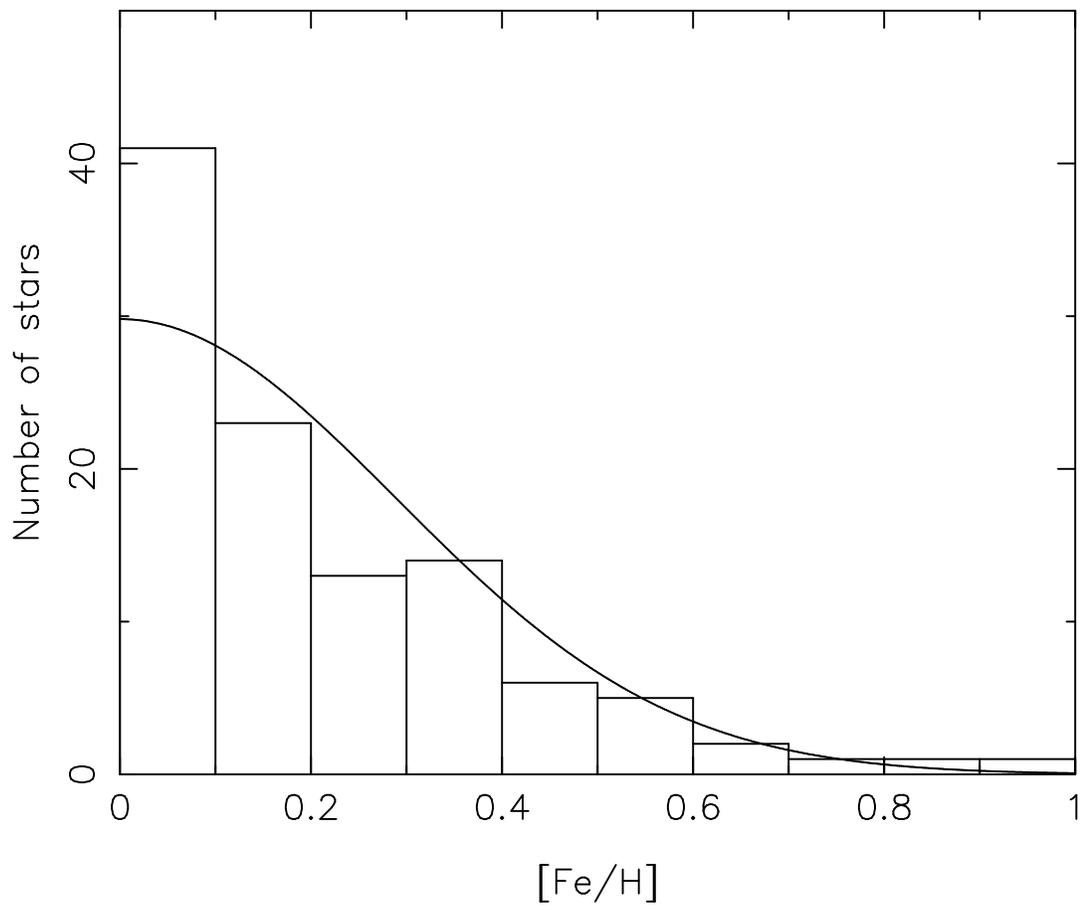

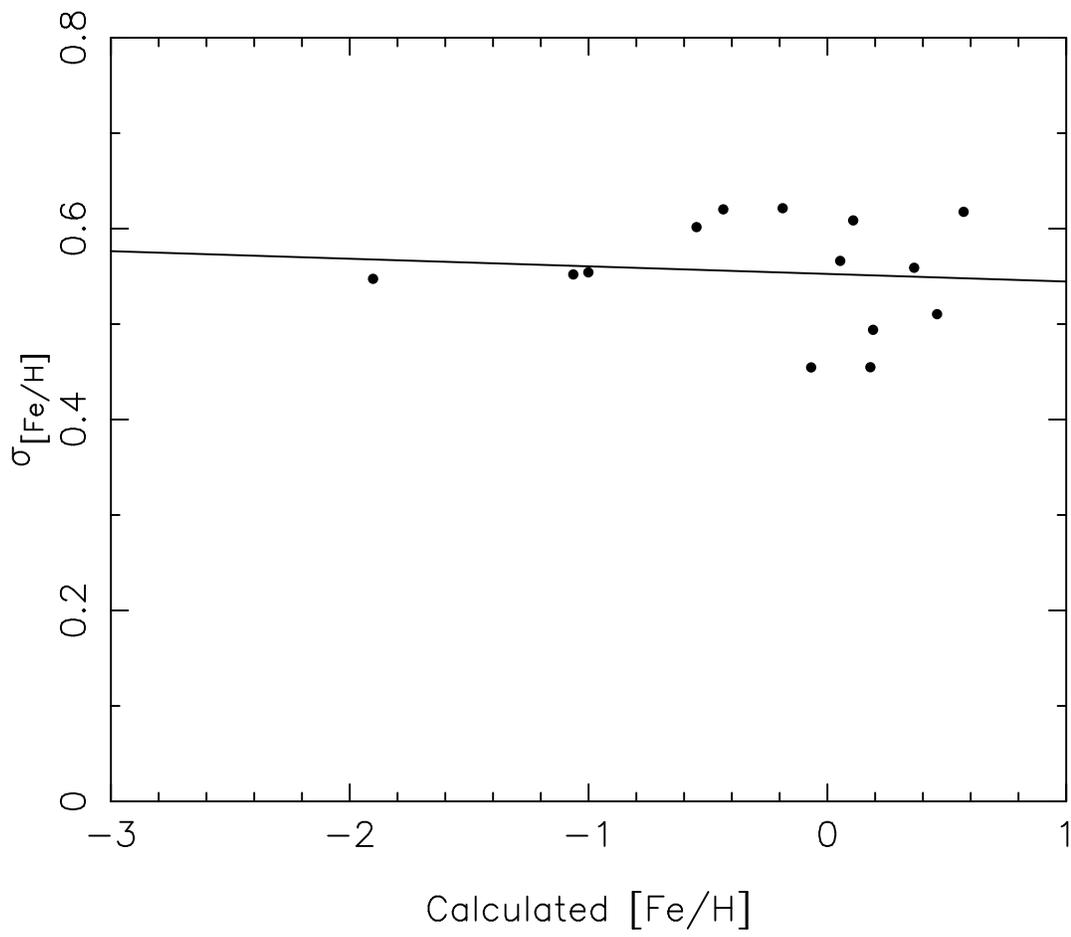

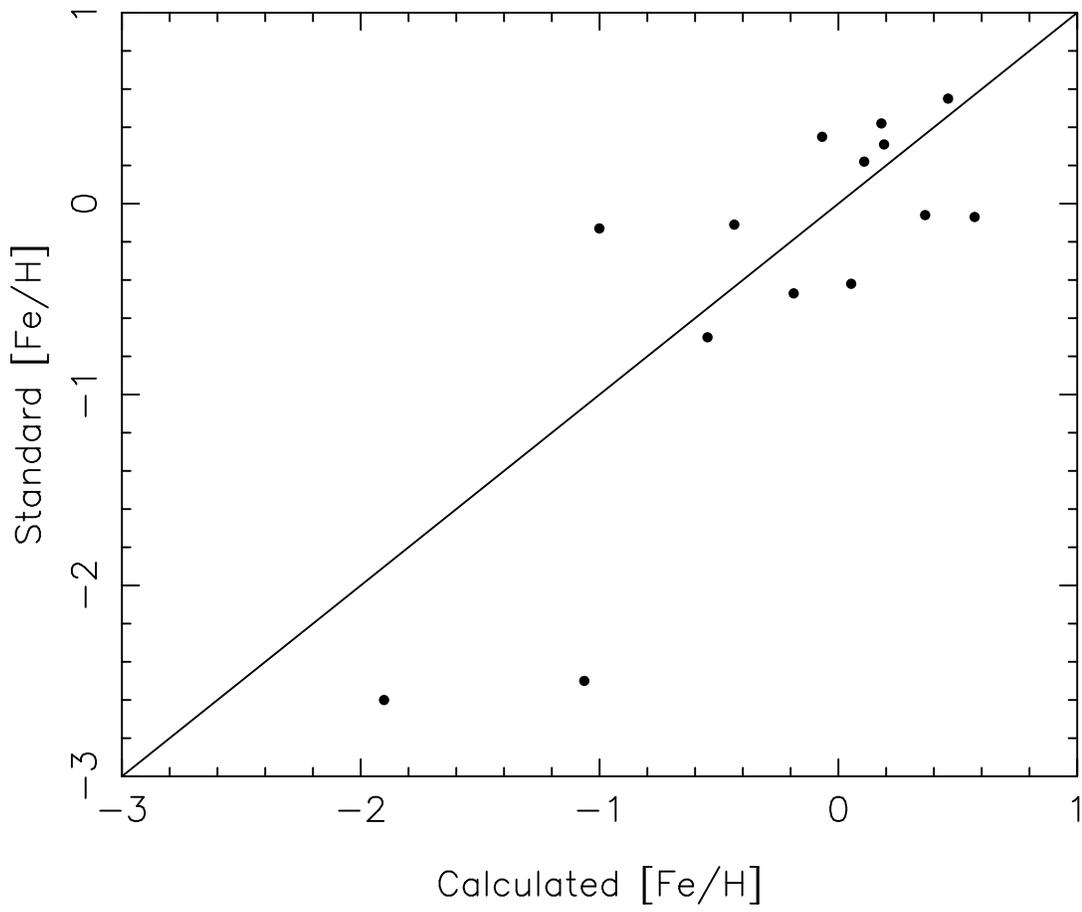

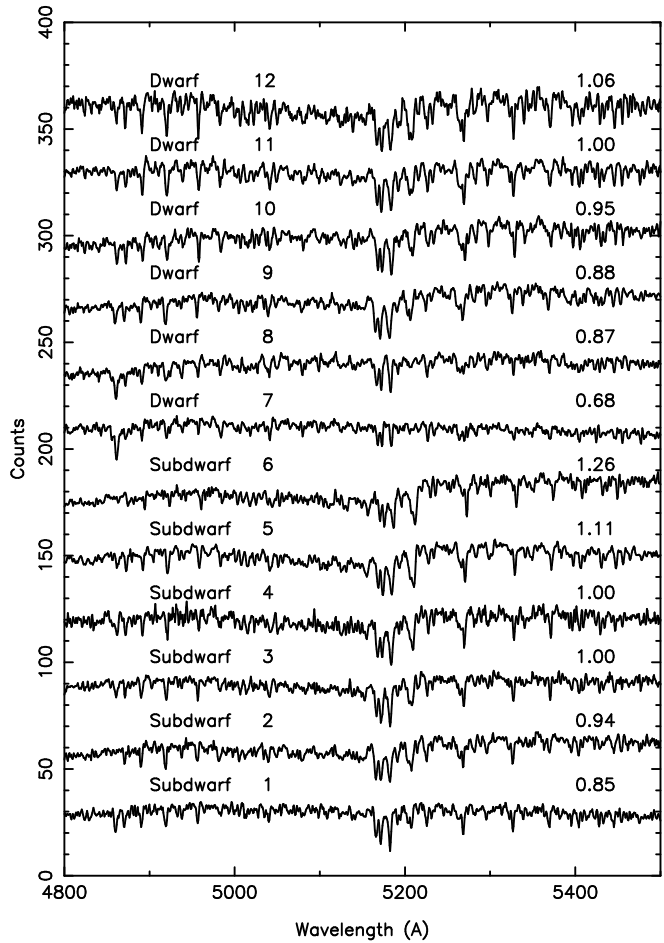
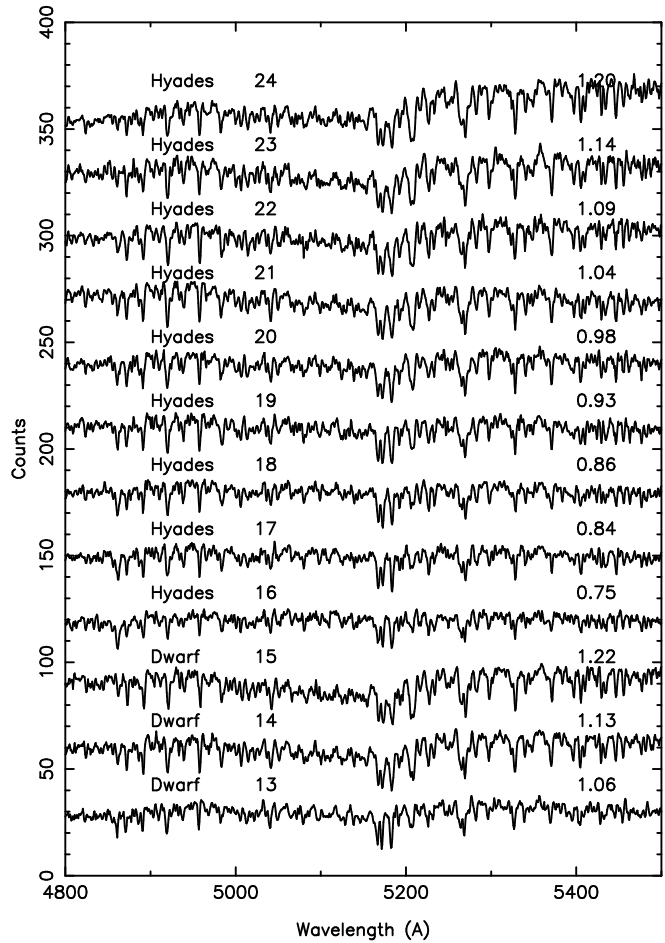
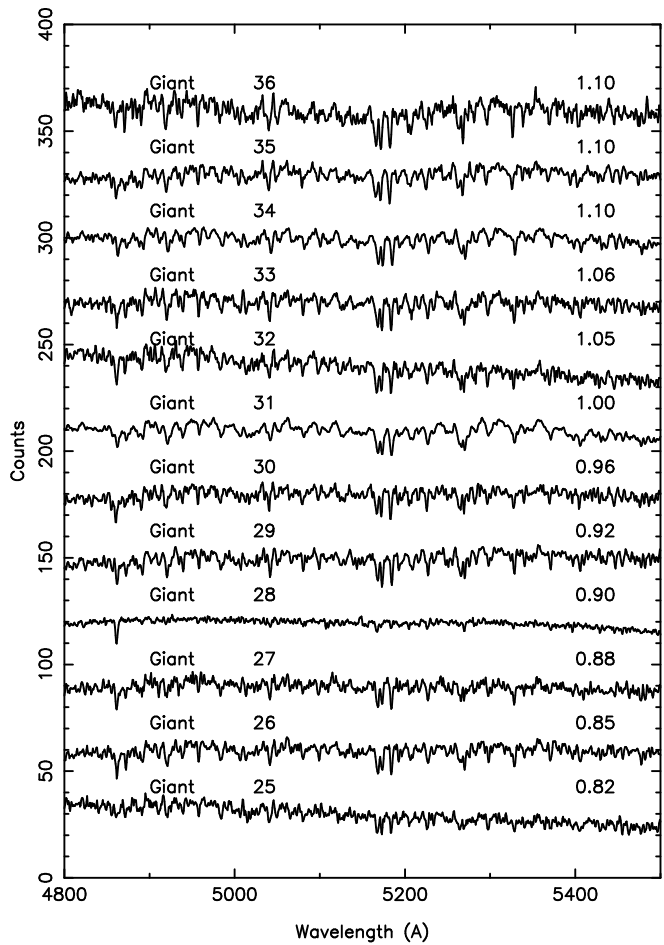
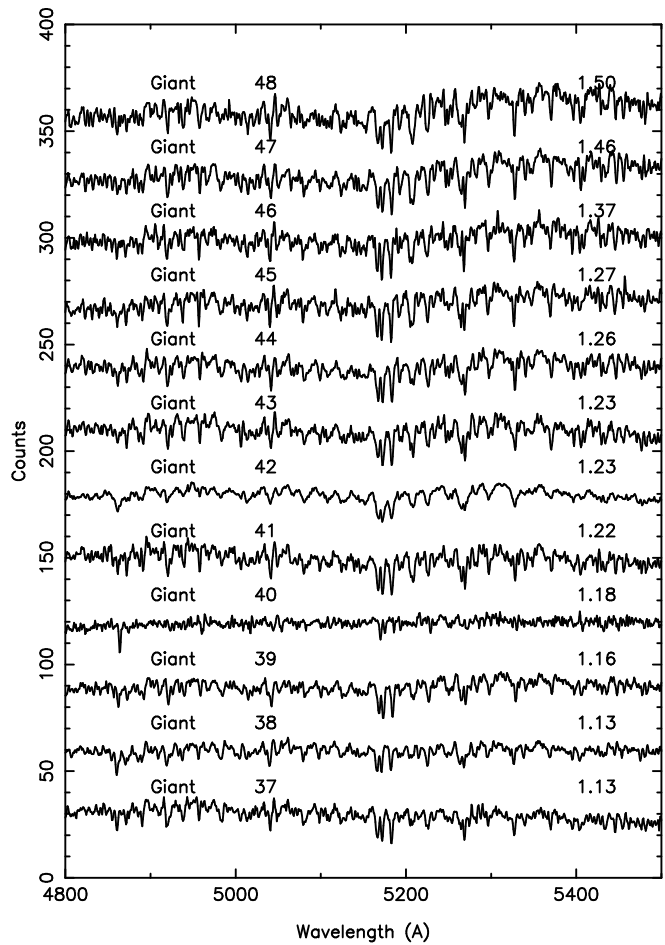

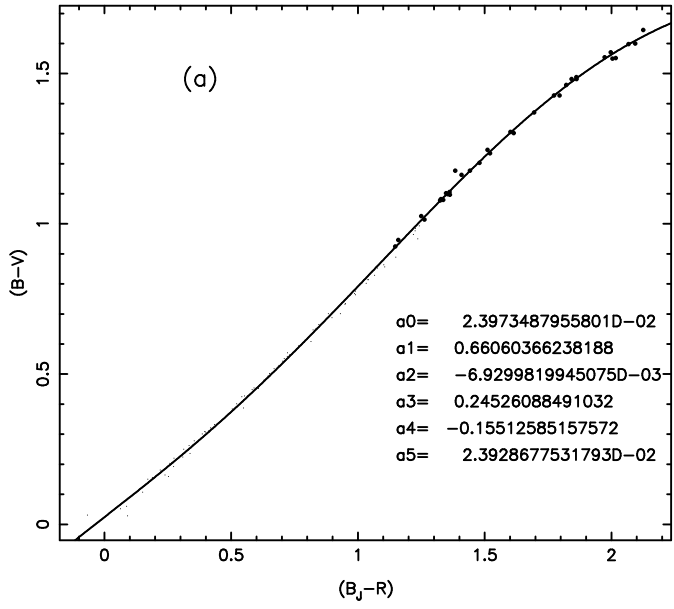
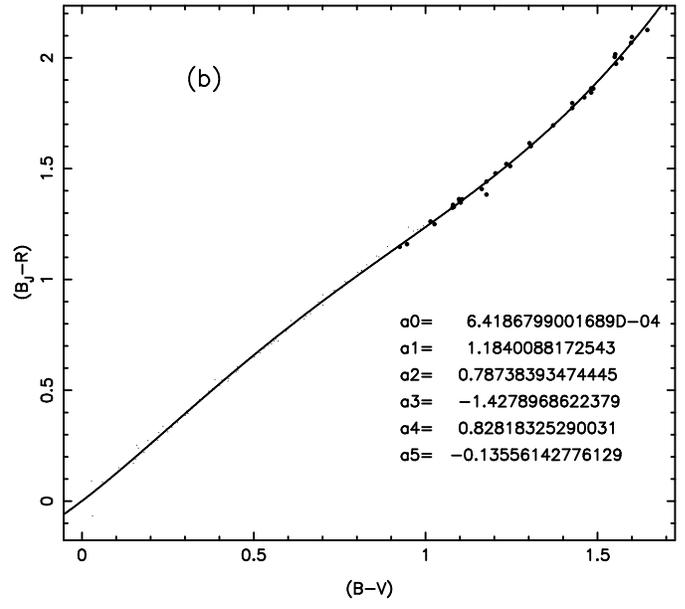
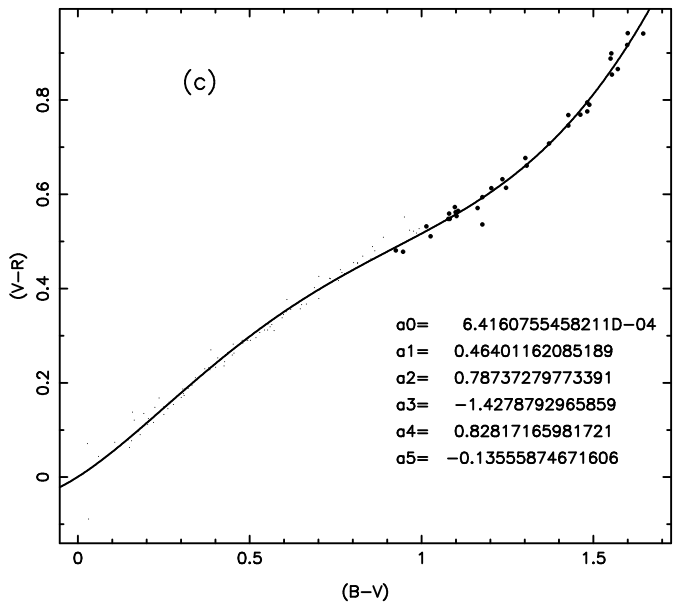
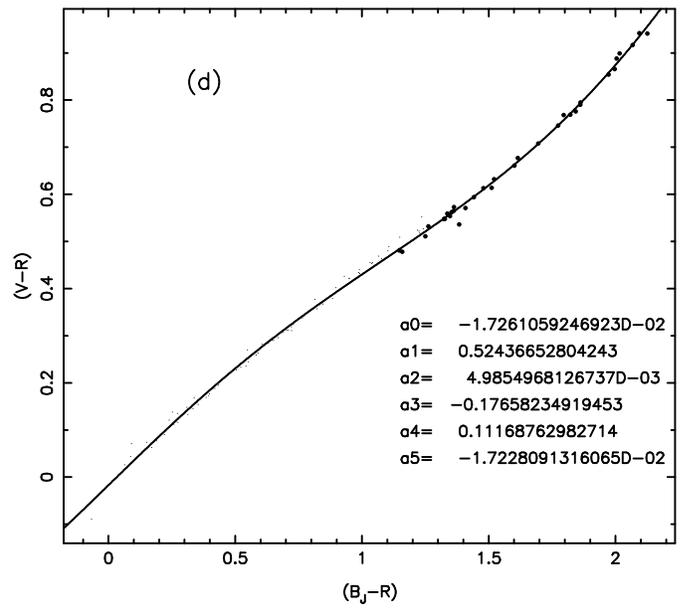

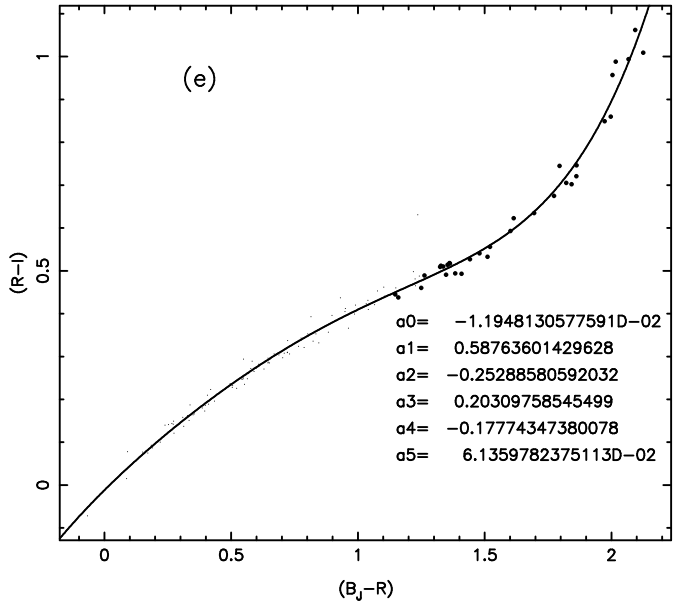
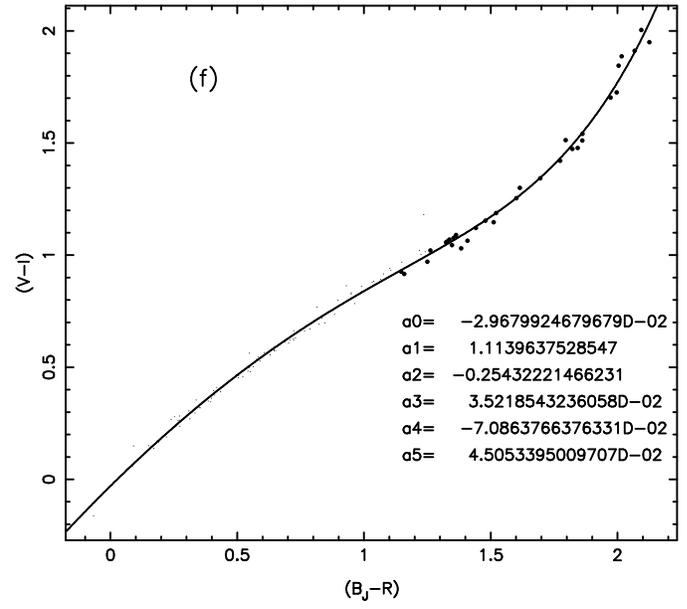
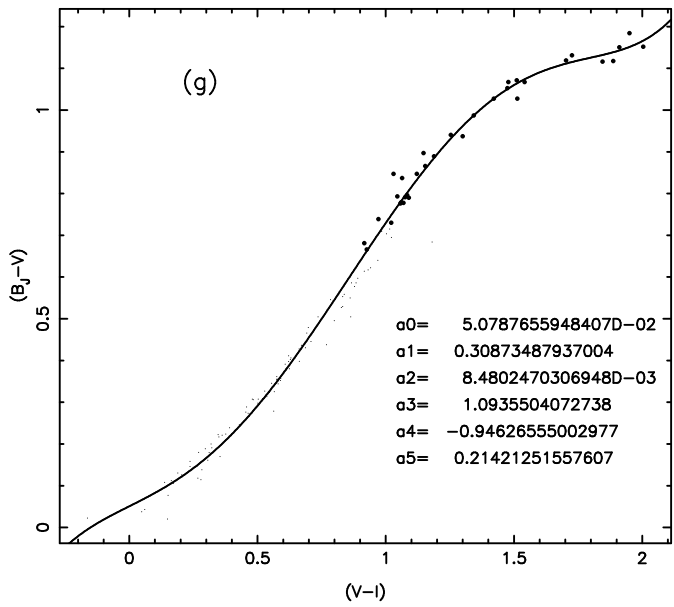
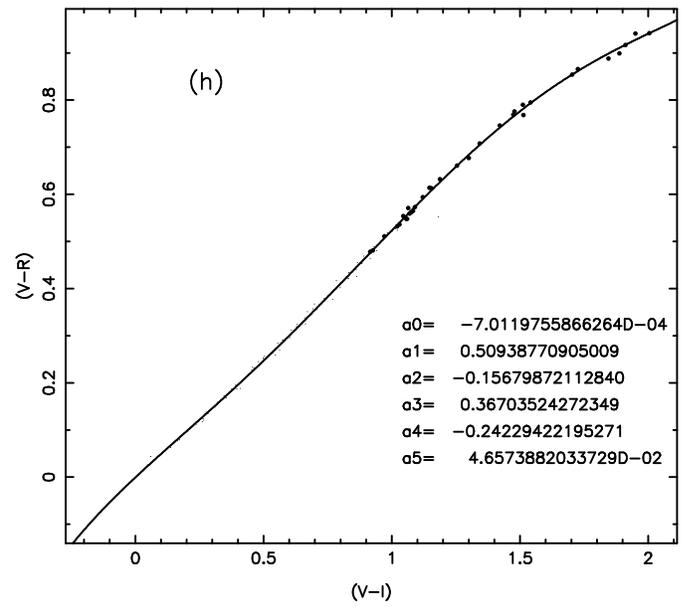